# Visible-Light Photocatalytic Degradation of Cresols using Sustainable 3D-Printed $Bi_4O_5I_2$-Hematite Scaffold

Akash Rawat[a,⊤], Raphael B. de Oliveira[b,c,⊤], Tapas Pal[d], Kleuton Antunes[c], Guilherme S. L. Fabris[c], Raphael M. Tromer[e], Marcelo L. Pereira Junior[b,f], Adarsh Singh[g], Ashok Kumar Gupta[g,*], Douglas S. Galvão[c,*], Chandra Sekhar Tiwary[h,*]

[a] *School of Environmental Science and Engineering, Indian Institute of Technology Kharagpur, Kharagpur 721302, India.*

[b] *Materials Science and Nano Engineering Department, Rice University, Houston, TX, 77005, USA.*

[c] *Applied Physics Department and Center for Computational Engineering & Sciences, State University of Campinas, Campinas, São Paulo 13083-970, Brazil.*

[d] *School of Nanoscience and Technology, Indian Institute of Technology Kharagpur, Kharagpur 721302, India.*

[e] *University of Brasília, Institute of Physics, Brasilia 70910900, Federal District, Brazil.*

[f] *University of Brasília, College of Technology, Department of Electrical Engineering, Brasilia 70910900, Federal District, Brazil.*

[g] *Environmental Engineering Division, Department of Civil Engineering, Indian Institute of Technology Kharagpur, Kharagpur - 721302, India.*

[h] *Metallurgical and Materials Engineering, Indian Institute of Technology Kharagpur, Kharagpur - 721302, India.*

*Corresponding authors (Email: agupta@civil.iitkgp.ac.in; chandra.tiwary@metal.iitkgp.ac.in; galvao@ifi.unicamp.br)

⊤*Equal contribution*

**Abstract**

In photocatalysis, the reusability limit of catalysts can contribute to secondary pollution, posing ecological risks. Addressing this, the present study explores the integration of additive manufacturing with photocatalysis by decorating $Bi_4O_5I_2$ onto a 3D-printed hematite scaffold ($Bi_4O_5I_2$@3DH) for the degradation of cresols. The 3D-printed hematite grid, fabricated via direct ink writing, exhibited excellent rheological behavior ($\tau_y$ = 24 Pa), allowing precise shape retention. The sintered $Bi_4O_5I_2$ was subsequently immobilized via a facile dip-coating method. Under optimized conditions, the composite achieved 99.78% degradation of 20 mg/L $p$-cresol within 240 min of irradiation. Notably, hematite served as a porous substrate and contributed to photocatalytic activity. Density functional theory simulations with Hubbard correction (DFT+$U$), indicated an interfacial charge transfer of approximately -0.9 electrons from hematite to $Bi_4O_5I_2$, confirming a S-scheme heterojunction between hematite and $Bi_4O_5I_2$ semiconductors, validating experimental observations. The composite demonstrated strong performance across varied water matrices and in the presence of other cresol isomers. It also retained 84.28% degradation efficiency after 10 cycles, with negligible catalyst leaching. Furthermore, *in vitro* and *in silico* ecotoxicity analyses revealed reduced toxicity of the degradation products. The current work presents a novel and scalable strategy, advancing the use of earth-abundant hematite minerals in sustainable environmental remediation.

**Synopsis**

We present the first 3D-printed hematite-based photocatalyst, decorated with $Bi_4O_5I_2$, enabling rapid cresol degradation, minimal catalyst loss, and reduced byproduct toxicity.

Supported by DFT simulation, it offers a scalable and eco-friendly wastewater remediation solution.

## 1. Introduction

The economic expansion of a nation is driven by prompt industrialization and urbanization. However, this progress is often accompanied by environmental challenges,[1–3] such as water pollution caused by the discharge of various toxic chemicals into the surface and subsurface water environments.[4,5] One such class of frequently encountered chemicals is cresol, typically originating from industries including coal gasification, coal conversion,[6] chemical production, medicines, petrochemicals, herbicides, and fungicides.[7] Cresols are highly soluble in water, with solubility ranging from 1.9 to 2.5 g/100 mL at room temperature,[8] and are microbial growth inhibitors.[9] Thus, they persist in wastewater treatment plant effluent and receiving water bodies.[10] These xenobiotic chemicals pose a risk to microbes, and their chronic exposure, even at low concentrations, can have pernicious effects on human eyes, skin, mental well-being, and the respiratory system.[11,12] According to the World Health Organization (WHO), the permissible concentration of *p*-cresol in potable water is 1.0 µg/L.[13] Furthermore, the Federal Water Pollution Control Act classifies cresols as hazardous substances, and the Clean Water Act Amendments (1977 and 1978) regulate their release.[14] Consequently, various treatment technologies, such as photocatalysis,[15,16] adsorption,[17,18] sonolysis,[19] and biodegradation[20] have been explored to mitigate these toxic pollutants.

Among these treatments, photocatalysis has emerged as a promising solution,[21,22] owing to its ability to harness solar energy to mineralize a wide range of aqueous pollutants.[23,24] Nevertheless, despite the several advantages of photocatalysis, its practical application is hindered by post-treatment challenges, including the low recovery yield of catalysts and the risk of secondary pollution from their leaching in conventional slurry-based systems.[25] To overcome these challenges, Significant efforts have been made towards stabilizing the catalysts onto solid supports, including polymers, clay, glass, calcium alginate, ceramics, activated carbon, and metals.[25–33] Although these catalyst carriers reduced the risk of nanoparticle leaching into the treated water, their inert and non-porous nature likely restricted mass transfer and active site accessibility, limiting overall photocatalytic efficiency. Contrastingly, photoactive and porous supports can offer synergistic interactions with the catalyst, enhancing both mass transport and exposure of active sites to reactants, thus improving photocatalytic performance. In this context, additive manufacturing is highly suitable and an attractive technique for fabricating catalyst support.[34] It allows the design of structures from basic to complex geometries, offering a higher area-to-volume ratio at the macroscopic level than conventional supports.[35,36] Among various 3D printing methods,[37] direct ink writing (DIW) is typically preferred due to its cost-effectiveness, facile operation, and ability to print a wide range of materials, including metals, ceramics, polymers, and glass.[38–40] Essentially, DIW inks require thixotropic properties to ensure shear-thinning and high viscosity, allowing them to withstand their shape post-extrusion.[40,41] The concentration of particle additives significantly influences the ink's viscosity and shear-thinning behaviour,[42] and these additives are typically removed to generate porous 3D printed structures,[43] where the additive concentration directly governs porosity.[44]

Hematite has abundant reserves and notable capabilities for visible light absorption.[25,45] However, its effectiveness in photocatalysis is constrained by low electron transport efficiency and a tendency for rapid charge carrier ($e^-/h^+$) recombination.[46] Unlike the inert supports mentioned above, when a semiconductor material such as hematite is decorated with a visible-light-driven photocatalyst such as $Bi_4O_5I_2$, its inherent optical properties can be controlled more effectively.[25] Given this, the present investigation focuses on the decoration of $Bi_4O_5I_2$ onto 3D-printed hematite scaffolds (3DH) through an iterative dipping and drying methodology (referred to as $Bi_4O_5I_2$@3DH). Moreover, to gain further insights into the physical mechanisms of the interaction between $Bi_4O_5I_2$ and the 3DH we also carried out *ab initio* simulations. Initially, the factors affecting photocatalytic degradation, such as the amount of catalyst coated (L), infill density ratio (IDR) of the scaffold, initial pollutant concentration ($C_0$), and pH values, are optimised through the degradation of model pollutant *p*-cresol. Further, the optimized $Bi_4O_5I_2$@3DH structure was thoroughly characterized to determine the physico-chemical and electronic properties that eventually influence the photocatalytic process. Subsequently, the real field applicability of $Bi_4O_5I_2$@3DH was assessed in various water matrices. Furthermore, post-treatment analysis of treated aqueous *p*-cresol, including anticipation of degradation products and their bio-toxicity, was evaluated using analytical methods and Quantitative Structure-Activity Relationship (QSAR) techniques.

## 2. Experimental section

### 2.1. Materials and chemical reagents

All chemical reagents provided in the S1 section of the ESI were used without further purification. However, the hematite ore collected from the Department of Mining, IIT Kharagpur, was washed thoroughly using DI water.

### 2.2. Instrumentation and characterization

The instruments used for characterization are listed in Section S2 of the ESI.

### 2.3. Fabrication of $Bi_4O_5I_2$@3DH

The $Bi_4O_5I_2$@3DH was synthesized through a three-stage process involving scaffold fabrication, $Bi_4O_5I_2$ synthesis, and their immobilization, as illustrated in Fig. 1 and detailed below:

*2.3.1. 3D printing of hematite scaffold (3DH)*

The ink was initially prepared by dissolving 5% (g/mL) carboxymethyl cellulose (CMC) in DI water, wherein CMC served as a binder. Although CMC is inherently hydrophilic due to its carboxymethyl groups, the solution was stirred at 500 rpm on a hot plate at 90 °C to ensure homogeneous dispersion. Finely ground hematite powder was then incorporated into the solution in an appropriate proportion to achieve the desired thixotropic behavior suitable for DIW. The resulting ink was utilized to fabricate hematite grids with varying IDR (25%, 50%, and 75%) using a DIW printer (Hyrel Engine HR, USA). Prior to printing, a digital model was created using the REPETREL (version 42.305) slicing software. After several design iterations, the optimized printing parameters were finalized and are presented in Table S1. These included a slice height of 0.6 mm, a nozzle diameter of 1.2 mm, a printing speed of 1 mm/s, and a grid infill pattern, with all operations conducted at room temperature (Video S1). The as-printed green hematite structures were air-dried at room temperature for a day, then sintered in a muffle furnace at 950 °C for 6 h with a heating ramp rate of 5 °C/min. The sintered structures were allowed to cool naturally to room temperature and are referred to as 3D-printed hematite scaffolds (3DHs).

*2.3.2. Synthesis of $Bi_4O_5I_2$ photocatalyst*

The preparation of the $Bi_4O_5I_2$ photocatalyst is briefly outlined in section S3 of the ESI.

*2.3.3. Fabrication of $Bi_4O_5I_2$@3DH*

In a 100 mL beaker, 1% (w/v) coating suspension was prepared by adding 200 mg of $Bi_4O_5I_2$ photocatalyst in 20 mL DI water, followed by ultrasonication for 30 min to attain uniform dispersion. Afterwards, the as-fabricated 3DH was immersed in suspension for 30 seconds and then gradually withdrawn, constituting a dip-coating cycle (N). Three separate 3DH samples were subjected to 5, 10, and 20 dip cycles to optimize photocatalyst loading. Notably, after each cycle, the coated 3DH was dried at 80 °C for 2 h in a hot air oven.

However, following the final dip, the $Bi_4O_5I_2$ decorated 3DH composites were dried for a day at 65 °C and are hereafter referred to as $Bi_4O_5I_2$@3DH scaffold. Fig. 2a and b present the $Bi_4O_5I_2$@3DH samples with varied N and IDR values.

## 2.4. Experimental set-up and procedure

The photocatalytic reactor (PCR), illustrated in Fig. 2c, primarily comprises a 50-watt LED light source, a 200 mL jacketed beaker, a stainless-steel frame for sample holding, and a magnetic stirrer. Apart from the inclusion of the steel frame, the reactor configuration remains consistent with that employed in our previous studies.[4,25] Further details of the experimental setup with a photograph of PCR are provided in Section S4 and Fig. S1 of the ESI, and a real-time video of the PCR during photocatalysis is available as Video S2.

The experimental procedure started with dark adsorption studies to determine the time required to attain adsorption-desorption equilibrium, using 20 mg/L *p*-cresol solution and fabricated samples. Accordingly, all subsequent photocatalytic experiments adopted the optimized dark phase duration. A one-variable-at-a-time (OVAT) approach was employed to systematically evaluate key operational parameters, starting with the optimization of N and IDR values.[4] Followed by investigations into the effects of $C_0$, solution pH, and cresol derivatives. The residual concentrations of *p*-cresol were quantified using high-performance liquid chromatography (HPLC) or total organic carbon (TOC) analysis. The detailed experimental procedure is provided in Section S5 of the ESI.

## 2.5. DFT calculations

The methodology used for the DFT simulations is provided in Section S6, and the optimized structures illustrating bulk hematite, $Bi_4O_5I_2$, and the hematite (001) surface with the 4×4 supercell are depicted in Fig. S2.

## 2.6. Ecotoxicity evaluation

The colony-forming unit (CFU) method was employed to assess the bio-toxicity of photocatalytically treated water against *E. coli*, as per studies.[4,25,47] In addition, *in silico* toxicity prediction was conducted using the Toxicity Estimation Software Tool (T.E.S.T.),[48] which applies QSAR models to estimate the potential toxicity of the intermediates formed during and after the photocatalytic degradation of *p*-cresol.

## 3. Results and discussion

### 3.1. Rheological analysis of DIW ink

Fig. S3a shows that shear stress increases proportionally with shear rate. Fig. S3b further illustrates that shear viscosity decreases with increasing shear rate, indicating that the ink exhibits shear-thinning behavior. This property enables the ink to flow smoothly through the nozzle even under low extrusion pressure during DIW.[49] Shear thinning is a non-Newtonian characteristic, which is desirable for DIW applications.[40] When extrusion pressure is applied to the syringe, the ink flows toward the nozzle tip, where it encounters a high shear rate and reduced viscosity, thus facilitating improved printability. Upon exiting the nozzle, the ink experiences a lower shear rate and higher viscosity, which is crucial for maintaining the dimensional stability of the printed structure.[50,51]

Fig. S3c presents the plot of storage modulus (G') and loss modulus (G'') as a function of shear stress. The values of G' and G'' provide insight into the ink's solid-like and liquid-like behavior, respectively.[52] When G' > G'', the ink behaves like an elastic solid, which contributes to shape retention after printing. Conversely, when G' < G'', the ink exhibits liquid-like behavior and cannot maintain a defined shape upon deposition. At the point where G' = G'', the ink exhibits a gel-like behavior.[36,53] The intersection of the G' and G'' curves indicates the yield stress ($\tau_y$), which marks the transition from solid-like to liquid-like behavior.[54] In this study, the ink demonstrated a yield stress of 24 Pa. The ink shows excellent shape retention at lower shear stress levels, where G' > G''- even when multiple layers are printed. Beyond the yield point, G" exceeds G' (i.e., G' < G"), enabling the ink to flow easily through the nozzle during extrusion.[49,55]

### 3.2. Characterization of $Bi_4O_5I_2$@3DH

The field emission gun scanning electron microscope (FEG-SEM) images of the 3DH scaffold and $Bi_4O_5I_2$@3DH composite are depicted in Fig. 3a, demonstrating their morphological features. The FEG-SEM image of the bare 3DH scaffold (Fig. 3a(i)) reveals the emergence of macro-porosity, which could be attributed to thermal expansion/contraction during the sintering/cooling processes. The high-magnification image (Fig. 3a(ii)) shows micro and nanopores distributed over the 3DH surface (surrounded by hematite crystals), likely formed due to the decomposition of the CMC binder during sintering at a given temperature.[56] These micropores could act as potential anchoring sites for $Bi_4O_5I_2$ nanoparticles. As evident from Fig. 3a(iii), the $Bi_4O_5I_2$ coating reduces the apparent macro-porosity, while the micropores appear to be filled with $Bi_4O_5I_2$ nanoparticles, as shown in Fig. 3a(iv), reinforcing stronger integration.

The X-ray diffraction (XRD) patterns of $Bi_4O_5I_2$@3DH, $Bi_4O_5I_2$, and 3DH are presented in Fig. 3b. As displayed, the composite sample ($Bi_4O_5I_2$@3DH) exhibits distinct diffraction peaks attributable to both 3DH and $Bi_4O_5I_2$, confirming the successful deposition of $Bi_4O_5I_2$ nanoparticles onto the 3DH scaffold. Specifically, the diffraction peaks at 2θ values of 28.10°, 31.45°, 45.15°, 49.30°, and 54.40° correspond to the (−114), (020), (224), (−424), and (118) planes of $Bi_4O_5I_2$. Whereas peaks located at 33.08°, 35.60°, 49.24°, 54.00°, 62.40°, and 63.90° are assigned to the (012), (104), (110), (024), (116), (214), and (300) planes of hematite.[57,58] These patterns closely match the standard reference cards #96-810-4104 for monoclinic $Bi_4O_5I_2$ (a = 11.26 Å, b = 5.7 Å, c = 14.95 Å)[59,60] and #00-024-0072 for rhombohedral hematite (a = b = 5.038 Å, c = 13.72 Å).[61] Notably, a slight shift in the diffraction peaks towards lower 2θ values is observed in the composite, indicating possible lattice strain or interfacial interaction between the constituent materials, also noticed by other authors in their investigations.[4,25,62,63] Such structural coupling is anticipated to facilitate improved charge separation and thereby enhance the photocatalytic degradation of cresols.

In addition, our previous study has comprehensively established the optical properties, including ultraviolet-visible diffuse reflectance spectroscopy (UV-DRS), photoluminescence (PL) spectroscopy and ultraviolet photoemission spectroscopy (UPS), and electrochemical behavior such as electrochemical impedance spectroscopy (EIS) of hematite, $Bi_4O_5I_2$ and the hematite/$Bi_4O_5I_2$ composite.

### 3.3. Photocatalytic performance of the $Bi_4O_5I_2$@3DH scaffold

The synergetic interaction between adsorption and photocatalysis is extensively documented in the literature.[64] In view of this, a one-hour adsorption study was conducted under the given operational parameters (corresponding cresol isomer concentration: 20 mg/L; IDR: 50%; N: 5; pH: 6.5), shown in Fig. S4a-c. The adsorption of the model contaminant (*p*-cresol) was found to be less than 1% (Fig. S4a). In contrast, saturation in the adsorption of *o*-cresol (Fig. S4b) and *m*-cresol (Fig. S4c) was achieved within 30 min, with adsorption efficiencies reaching approximately 10.3%. This level of adsorption was substantial and could not be ignored. Hence, the initial 30 min of the PCD experiments were conducted in the dark to attain the adsorption-desorption equilibrium.

*3.3.1. Factors influencing the photocatalytic degradation of p-cresols*

*(i) Number of dip-cycles*

Optimizing the value N is critical, as it directly affects the amount of $Bi_4O_5I_2$ catalyst loaded onto the 3DH scaffold, an important parameter influencing photocatalytic activity. As indicated in the inset of Fig. 3c, the 5, 10, and 20 N values resulted in L values of approximately 67.1 ± 4.7 mg, 124.6 ± 4.2 mg, and 237.6 ± 5.3 mg, respectively. Their corresponding PCD efficiencies against *p*-cresol (20 mg/L) are illustrated in Fig. 3c, revealing that the composite with N = 10 achieved the highest PCD efficiency (99.78%), followed by N = 5 (79.52%) and N = 20 (71.6%). The lower activity at N = 5 is likely a result of insufficient catalyst loading, which limited light absorption necessary for efficient PCD.[65] In contrast, the decreased efficiency at N = 20 is attributed to excessive catalyst deposition, possibly leading to agglomeration,[66] blockage of active sites, and the formation of a thick layer that suppressed the active role of the 3DH scaffold.[65] The N = 10 condition provided an optimal balance, enabling synergistic interaction between $Bi_4O_5I_2$ and the 3DH structure. These findings highlight the importance of tuning catalyst loading on semiconductor substrates employed for dual functionalities, including synergistic interaction as an active support and an immobilising carrier. This ensures effective interaction between the photocatalyst and the scaffold.

*(ii) Infill density ratio of 3DH scaffold*

Fig. 3d displays the influence of the IDR of the 3DH scaffold on the PCD of 20 mg/L *p*-cresol. Scaffolds with 25%, 50%, and 75% IDR were fabricated, carrying corresponding average weights of 2.751 ± 0.129 g, 2.262 ± 0.264 g, and 5.513 ± 0.059 g. As anticipated and displayed in the inset of Fig. 3d, after the application of $Bi_4O_5I_2$ coating (N = 10), the respective measured weights of $Bi_4O_5I_2$@3DH composites increased to 2.783 ± 0.131 g (L = 32.3 ± 2.2 mg), 3.386 ± 0.263 g (L = 123.6 ± 3.6 mg), and 5.680 ± 0.054 g (L = 167.5 ± 4.5 mg). Furthermore, under 4 h of LED irradiation, the observed PCD efficiencies for the IDR scaffolds were 68% (IDR = 25%), 99.78% (IDR = 50%), and 93.9% (IDR = 75%). Evidently, 25% IDR demonstrated the lowest efficiency, which can be attributed to inadequate surface area for the immobilization of the catalyst. On the other hand, the 75% IDR scaffold exhibited the highest catalyst loading; however, its PCD performance was slightly inferior to that of the 50% IDR. Such disagreement is likely attributed to the increased structural density, which may prevent light penetration. This tendency is consistent with results published by Claes et al.,[67] whereby fewer structural features in packed bed reactors increased catalyst density but imposed more light-scattering limits, hence lowering overall efficiency. Likewise, in as-fabricated 3DH, the denser infill possibly triggered internal light attenuation, therefore restricting the photocatalyst's efficacy. The 50% IDR scaffold recorded the highest PCD efficiency (99.78%), due to an optimal balance of exposed surface area and catalyst distribution. Thereby being used in the subsequent investigations.

*(iii) Initial concentration of p-cresol*

The initial concentration of *p*-cresol ($C_0$) is a crucial parameter for evaluating the PCD potency of a $Bi_4O_5I_2$@3DH, particularly under constant irradiation time (4 h). In this regard, the activity of $Bi_4O_5I_2$@3DH was assessed across a range of $C_0$ values (1-50 mg/L), as shown in Fig. 3e. The results revealed a progressive decline in PCD efficiency with increasing $C_0$. Specifically, for concentrations up to 20 mg/L, nearly complete PCD (~100%) was achieved within 4 h reactions, identifying 20 mg/L as the threshold for the comprehensive pollutant breakdown under the given conditions. The subsequent decline in PCD (beyond 20 mg/L) can be ascribed to the limited availability of active sites on $Bi_4O_5I_2$@3DH,[68] which were insufficient to degrade higher pollutant concentrations.[4,25] Hence, the $C_0$ of 20 mg/L was used in the subsequent analysis.

*(iv) Initial pH of aqueous solution*

Typically, the pH of the aqueous medium plays a pivotal role in the PCD of pollutants, as it governs both pollutant adsorption and catalyst stability.[4,69] In this study, the pH of the *p*-cresol solution was varied from 2.5 to 12.5 to assess its impact on PCD efficiency (Fig. 3f).

Reasonably consistent with our previous study,[25] the highest PCD was attained at unadjusted pH 6.5 (99.78%) followed approximately 10% and 4% decline at pH 7.5 and 5, respectively. However, a marked drop in PCD can be perceived at extreme pH values (2.5 and 12.5), consistent with bismuth oxyiodide materials-based investigations.[4,25]

The zeta potential analysis from our previous work[25] revealed a point of zero charge (PZC) of 5.5 for the hematite/$Bi_4O_5I_2$ composite, indicating a positively charged surface below this pH and a negative charge above it. Meanwhile, the dissociation constant ($pK_a$) of *p*-cresol (~10.3)[70,71] dictates its speciation: the neutral form dominates below pH 10.3, transitioning to its anionic form beyond this value (Fig. S5a). Thus, the decline in PCD efficiency at pH > 10.3 can be attributed to electrostatic repulsion between the negatively charged catalyst surface and anionic *p*-cresol, which hinders adsorption. Additionally, excess $OH^-$ ions may scavenge photogenerated hydroxyl radicals ($^{\bullet}OH$), as reported by Malefane[72] and Xiong et al.,[73], thereby reducing the availability of reactive species. Moreover, the redox potential of $^{\bullet}OH$ radicals decreases under alkaline conditions, contributing to the diminished photocatalytic activity at mild alkalinity.[74,75] Therefore, pH 6.5 is optimal for PCD using $Bi_4O_5I_2$@3DH, with a recommended operational range between pH 5 and 10.

### 3.4. DFT-supported charge transfer mechanism in $Bi_4O_5I_2$@3DH heterojunction

Experimental results confirmed the formation of an S-scheme heterojunction between hematite and $Bi_4O_5I_2$,[25] but several properties at the interface are still challenging to determine using only analytical techniques. These include the precise nature of charge redistribution at the heterointerface, local variations in band alignment, and the influence of intrinsic defects on electronic behavior. Such factors are crucial for understanding the charge transfer path and establishing the internal electric field (IEF). Therefore, density functional theory (DFT) simulations were carried out to gain atomistic insight into the structural and electronic interactions within the $Bi_4O_5I_2$@3DH composite that underpin its photocatalytic performance.

**3.4.1 DFT simulation of the interaction between hematite and $Bi_4O_5I_2$**

After geometry optimization, the lattice parameters obtained for the bulk structures were: a = b = 5.17 Å, c = 13.86 Å, α = β = 90.0°, γ = 120.0° for hematite; and a = 10.93 Å, b = 5.64 Å, c = 14.63 Å, α = 90.00°, β = 98.26°, γ = 90.0° for $Bi_4O_5I_2$. These values are in good agreement with the experimental data, with deviations of approximately 3%,[25,76] confirming the accuracy of the computational method employed. For the hematite (001) surface, the optimized 4×4×1 supercell presents the following parameters: a = b = 21.04 Å, c = 56.68 Å, α = β = 90.0°, γ = 120.0°. Regarding their electronic properties, the values calculated for hematite and $Bi_4O_5I_2$ in their respective bulk forms are consistent with experimental data and recent literature, exhibiting deviations of less than 10% .[25,76–78] For the hematite (001) monolayer used to model the substrate, comparisons with theoretical and experimental data must consider the strong dependence of the electronic band gap on the slab thickness. For systems with dimensions similar to those employed in this study, the simulations reported in the literature show a comparable trend in band gap values.[79] Finally, for the heterojunction, our results reproduce the same trend observed experimentally, with the simulations indicating a reduction of approximately 67% in the electronic band gap. It is worth noting that such discrepancies are expected when using the PBE functional, which is known to underestimate band gap values.[80]

Based on the agreement between the calculated structural and electronic properties[25] and the available experimental data, we proceeded to model the deposition process of $Bi_4O_5I_2$ onto the hematite substrate. Fig. 4a shows the initial configuration of the system, with the hematite (001) substrate, periodic along the x and y directions, and the $Bi_4O_5I_2$ flake positioned at an initial height of approximately 3.5 Å. During the structural optimization, $Bi_4O_5I_2$ moves closer to the substrate, resulting in a vertical displacement of some Fe atoms within the hematite structure (Fig. 4b). When considering an isolated $Bi_4O_5I_2$ fragment, some atoms lack full valence coordination and tend to form bonds with atoms from the substrate. In the final optimized geometry, shown in Fig. 4c, the formation of Fe-O bonds at the interface is observed. Importantly, Fig. 4c shows a displacement of an Fe atom close to an I atom, but no bond formations were found.

In the context of investigating photolysis and PCD of cresols (p-cresol, o-cresol, and m-cresol), employing 3DH and $Bi_4O_5I_2$, as well as their surface-based combinations, we aimed to elucidate the electronic phenomena occurring at the semiconductor interface. These interfacial effects are directly linked to the efficiency observed in the photocatalytic processes, as discussed in the experimental section. In this regard, Fig. 4d illustrates the evolution of the electronic band structure of the heterostructure formed between hematite and $Bi_4O_5I_2$, highlighting two key stages: the isolated electronic structures and the interface formed upon contact. We see in Fig. 4d the structures of the two isolated semiconductors $Bi_4O_5I_2$ and hematite, both of which exhibit behavior typical of n-type semiconductors, with their Fermi levels located closer to the conduction band. This initial alignment is crucial to understanding the nature of the formed heterojunction. The electronic structure no longer reflects a classic p–n junction but instead indicates the formation of a heterojunction of the S-scheme type, as shown by the band bending and spatial charge redistribution at the interface. In the S-scheme model, photogenerated electrons in the conduction band of $Bi_4O_5I_2$ and holes in the valence band of hematite are preserved after recombination of the less energetic carriers across the interface. The resulting band alignment, with a narrowed interfacial band gap (~0.6 eV), enables efficient separation of charge carriers, crucial for enhanced photocatalytic performance.

The realignment of Fermi levels upon contact leads to a built-in potential (Vbi), facilitating the migration of charge carriers and forming a depletion region characterized by the S-scheme interface. This region is illustrated in Fig. 4d with shaded areas and charge symbols, indicating the recombination of lower-energy carriers and preservation of the more reactive ones on opposite sides of the junction.

We also investigated the influence of point defects (vacancies) in both materials. In general, the semiconducting character of the materials is preserved in the presence of oxygen or iodine vacancies, which are thermodynamically more likely to occur. However, the removal of a Fe atom from hematite introduces metallic states, resulting in a more significant perturbation of the crystal lattice. Nevertheless, such a defect is considered less probable in this type of nanomaterial. Additionally, we observed that the presence of vacancies can reduce the work function by up to 1.5 eV, reaching values near 2.5 eV. Although this reduction affects the interfacial potential barrier, it does not inhibit the formation of the S-scheme heterojunction. Our findings suggest that vacancy-free models provide a more accurate representation of the experimentally observed behavior. Finally, we analyzed the charge transfer process between the hematite substrate and the $Bi_4O_5I_2$ flake based on the charge density variation illustrated in Fig. 5. This variation provides a spatial representation of how the electrons are redistributed throughout the system. In the context of heterostructures, the charge density difference ($\Delta\rho$) is particularly significant, as it reveals the interfacial charge transfer between the system's components. Regions with $\Delta\rho > 0$, represented in blue, indicate electron accumulation and, thus, an increase in negative charge. In contrast, regions with $\Delta\rho < 0$, shown in red, reflect electron depletion and a reduction in negative charge.

In the (001) hematite monolayer, shown in Fig. 5a, $\Delta\rho$ is negative around the Fe atoms and positive around the O atoms. This behavior is consistent with the structural features of the monolayer. Upon removal of periodicity along the direction now occupied by the vacuum, some oxygen atoms are left with incomplete valence shells. This facilitates a degree of electron donation from neighboring Fe atoms while maintaining the material's structural integrity. In Fig. 5b, a predominant electron gain is observed on the oxygen atoms, which is expected since oxygen requires relatively few electrons to complete its valence shell, leading the structure to seek a new electronically stable configuration. At the interface, illustrated in Fig. 5c, the analysis focuses on regions where Fe-O bonds are formed. A net charge loss is observed on the Fe atoms, accompanied by a corresponding charge gain on atoms belonging to the $Bi_4O_5I_2$ flake. To quantify this charge transfer, we calculated the difference in the total charge of each atom composing the (001) hematite monolayer before and after contact with $Bi_4O_5I_2$. This procedure revealed a net loss of approximately –0.9 electrons from the hematite side. Applying the same method to $Bi_4O_5I_2$, we obtained a net gain of +0.9 electrons after contact with the hematite substrate.

Despite this charge transfer, the total net charge of the (001) monolayer/$Bi_4O_5I_2$ system remained overall neutral. This net transfer of approximately –0.9 electrons from hematite to $Bi_4O_5I_2$ supports the classification of both materials as n-type semiconductors, with hematite acting as the electron donor and $Bi_4O_5I_2$ as the electron acceptor. The resulting redistribution and recombination of charge carriers at the interface confirm the formation of an S-scheme heterojunction, providing an ideal environment for charge separation and efficient photocatalytic activity, especially in reactions involving cresol degradation.

Importantly, DFT calculations are consistent with our previous study, demonstrating the flow of charge carriers across the hematite/Bi4O5I2 heterojunction.[25] Accordingly, a schematic representation of the validated S-scheme mechanism is provided in Fig. 6a, illustrating the induction of an IEF directed from hematite (RP: reduction photocatalyst) to $Bi_4O_5I_2$ (OP: oxidation photocatalyst).[81,82] However, charge carriers are photoexcited and spatially separated within their respective semiconductors upon light irradiation. The photogenerated holes in the valence band of OP participate in the oxidation of water molecules to produce hydroxyl radicals ($^\bullet$OH), while the electrons in the conduction band of RP reduce oxygen molecules to form superoxide radicals ($^\bullet$O$_2^-$). These reactive oxygen species (ROS) then work synergistically to degrade cresols into intermediates, eventually mineralizing them into $CO_2$ and $H_2O$. Meanwhile, low-energy charge carriers recombine at the heterointerface,[83] enhancing the selectivity and efficiency of the PCD process. In addition, electron paramagnetic resonance (EPR) spectroscopy was employed to confirm the generation of these radicals. As shown in Fig. 6b, the characteristic DMPO-$^\bullet$O$_2^-$ and DMPO-$^\bullet$OH adduct signals were detected after 20 min of visible light irradiation, thereby validating the radical formation during photocatalysis.[25]

*3.6. In vitro* **and** *in silico* **bio-toxicity assessment**

As evident from Fig. S6, the 4 h light irradiation on Bi$_4$O$_5$I$_2$@3DH composite led to the partial mineralization of aqueous *p*-cresol (TOC removal efficiency, i.e., TRE 74.63%), resulting in the formation of various intermediate and degradation products (IDPs). In order to identify these IDPs, four representative samples were extracted from the photocatalytic reactor at t = -30, 60, 120, and 240 min. Subsequently, the extracted samples were analyzed using LC-MS/MS, and the IDPs were identified based on their *m/z* ratios in the mass spectra shown in Fig. S7, which were further corroborated by literature reports.[83,84] Accordingly, the plausible degradation pathways of *p*-cresol were illustrated in Fig. 7a, showing the transformation of the parent compounds into various IDPs denoted as TP-n (where *n* = 0, 1, 2, …), followed by their eventual mineralization into $CO_2$ and $H_2O$, as explained in section S7 of ESI.

In the *in vitro* test, explicit CFU disparity was observed among the three Petri dishes (Fig. 7b); the DI water (control) exhibited the highest CFUs, followed by a photocatalytically treated sample and lastly, the untreated sample, shown in Fig. 7b(i-iii). These observations demonstrate the reduction in bio-toxicity after PCD, and the resemblance of CFUs between control and treated samples advocates the reduced bio-toxicity induced by the IDPs.

On the other hand, to conduct the *in silico* ecotoxicity assessment, a QSAR model-based TEST software[48] was used, which estimated crucial ecotoxicological endpoints based on molecular structure. These endpoints include acute aquatic toxicity (e.g., $LC_{50}$ for *Daphnia magna* and *fathead minnow*), binary endpoints (mutagenicity and developmental toxicity, referred to as DT), and bioaccumulation factor (BCF). The predicted results for the IDPs of p-cresol are summarized in Table S2, which includes their chemical structures, $LC_{50}$ values, and corresponding ecotoxicological profiles. Based on their $LC_{50}$ values, the *p*-cresol is categorized as "harmful" according to the Globally Harmonized System of Classification and Labelling of Chemicals (GHS).[85] However, *p*-cresol showed negative DT, suggesting no significant DT. These findings are detailed in Table S2. Furthermore, the IDPs displayed varied toxicity profiles, with $LC_{50}$ and binary endpoints delineated in Fig. 7c and 7d, respectively. Consistent with the CFU observations, the toxicity parameters predicted using TEST also indicated a reduction in the ecotoxicity associated with the various IDPs.

### 3.7. Practical significance of $Bi_4O_5I_2$@3DH composite

#### 3.7.1. Application of $Bi_4O_5I_2$@3DH in various water matrices

To enable commercial applications, a catalyst must be robust enough to withstand the complex and often unfavorable conditions of various real water matrices, such as tap water, pond water, and secondary treatment effluent (STE) from a sewage treatment plant (STP). Table S3 of the ESI provides the physico-chemical characteristics of these matrices. The PCD of *p*-cresol using $Bi_4O_5I_2$@3DH was evaluated under the optimized operational conditions described in Section 3.2.1. As shown in Fig. 8a, the highest PCD was achieved in tap water (92.72%), followed by pond water (68.26%) and STE (58.21%). Compared to the control (DI water), the reduced photocatalytic activity in real water matrices is attributed to the presence of co-existing ions and organic/inorganic constituents, which interfere with the degradation process through radical scavenging (e.g., by anions), light attenuation (due to turbidity and chemical oxygen demand, COD), and deactivation of active sites (by COD and nitrate). This trend in PCD efficiency (tap water > pond water > STE) is consistent with previous photocatalytic studies on real water matrices and corresponds to the increasing levels of interfering substances reported in Table S3, with the highest in STE and the lowest in tap water.[4,25,86] These findings highlight the applicability of the $Bi_4O_5I_2$@3DH composite across diverse water matrices.

*3.7.2. Simultaneous removal of cresols*

In addition to co-existing ions, co-existing organic contaminants represent another critical challenge that catalysts encounter in the PCD process's real-world applications. In this regard, the PCD of a mixture of cresol isomers (*p*-cresol, *m*-cresol, and *o*-cresol) was examined under the optimized conditions using the $Bi_4O_5I_2$@3DH scaffold. The initial concentrations of each isomer in the mixture varied from 1 mg/L to 20 mg/L. Moreover, the individual PCD performance of *p*-cresol, *m*-cresol, and *o*-cresol is presented in Fig. 8b-d of the ESI.

Of note, owing to the similar retention times of *p*-, *o*-, and *m*-cresol (provided in Section S5), the overlapping HPLC peaks were observed in samples containing the cresol mixture. This overlap could lead to an inaccurate assessment of individual degradation efficiencies. Therefore, the simultaneous degradation of the isomeric mixture was evaluated in terms of TRE, as shown in Fig. 8e. The TRE decreased from 98.74% to 56.17% as the initial concentration of the cresol isomers increased from 1 mg/L to 20 mg/L, indicating a concentration-dependent reduction in PCD efficiency. However, at 20 mg/L, the TREs for the individual isomers (Fig. S6) were 74.63% (*p*-cresol), 72.51% (*o*-cresol), and 71.88% (*m*-cresol), suggesting that the presence of multiple isomers significantly affected the overall photocatalytic performance of $Bi_4O_5I_2$@3DH.

*3.7.3. Reusability and stability test*

Although immobilization facilitates catalyst recovery from aqueous systems, its long-term stability and reusability must also be carefully evaluated to ensure sustained performance in practical applications. In this context, the $Bi_4O_5I_2$@3DH structure was subjected to ten consecutive PCD cycles of *p*-cresol under optimized conditions. As shown in Fig. 8f, only a decrease of 15.72% in degradation efficiency was observed, which may be attributed to the gradual loss of photocatalyst and active sites during reuse.[87] However, the inductively coupled plasma mass spectrometry (ICP-MS) analysis confirmed that any leaching of active components was below the instrument's detection limit, indicating good chemical stability. Importantly, the photocatalyst was reused in successive cycles without any further cleaning or regeneration. The synthesized $Bi_4O_5I_2$@3DH exhibited good structural stability and reusability, underlining its potential for use in continuous-flow reactors for real-field applications.

In addition, Table S4 presents a contrasting overview of the photocatalytic degradation performance of $Bi_4O_5I_2$@3DH scaffold and other reported immobilized photocatalyst systems targeting PCs. Notably, the present study distinguishes itself by achieving near-complete degradation of cresol isomers under visible-light irradiation using a 50 W LED source.

## 4. Conclusions

In summary, this study presents a $Bi_4O_5I_2$-decorated 3D-printed hematite scaffold ($Bi_4O_5I_2$@3DH) for the efficient PCD of cresol isomers. Unlike inert supports (e.g., glass or ceramic), a photoactive hematite scaffold was designed to contribute synergistically to the overall photocatalytic activity through $e^-/h^+$ transfer at the $Bi_4O_5I_2$ - hematite interface. Furthermore, the XRD analysis confirmed the interaction between $Bi_4O_5I_2$ and hematite, while FEG-SEM revealed micro- and macro-porosity in hematite that facilitated catalyst anchoring.

An earth-abundant hematite 3D-printed scaffold printed using DIW, exhibited excellent rheological behavior ($\tau_Y$ = 24 Pa), which ensured shape retention during multilayer printing under $G'' > G'$ conditions, enabling smooth extrusion and structural integrity. Following drying and sintering, $Bi_4O_5I_2$ was decorated onto the hematite scaffold by dip-coating the sintered scaffold in a 1% (w/v) $Bi_4O_5I_2$ suspension. After drying, the coated scaffold was systematically evaluated against the PCD of the model contaminant (*p*-cresol) under visible light irradiation. Initially, the infill density of the hematite scaffold was optimized, followed by initial *p*-cresol concentration and solution pH using the OVAT approach. The results revealed that a 50% infill density with ten dip-coating cycles yielded the highest PCD efficiency of 99.78% against *p*-cresol (20 mg/L) under 240 min of LED irradiation, with minimal adsorption of less than 5%. The EPR results showed characteristic DMPO-$^•O_2^-$ and DMPO-$^•$OH adducts, demonstrating the active participation of $^•O_2^-$ and $^•$OH radicals in the photocatalysis. Moreover, to better understand the electronic structure, charge transfer pathways, and active sites driving the observed photocatalytic performance, the DFT+*U* simulations confirm an interfacial charge transfer of approximately -0.9 electrons from hematite to $Bi_4O_5I_2$, supporting the formation of S-scheme heterojunction. The interfacial charge separation facilitated the generation of ROS ($^•$OH and $^•O_2^-$), which typically drives the PCD. Furthermore, partial mineralization of *p*-cresol (74.63%) led to the formation of IDPs, which were detected by LC-MS/MS. Ecotoxicity assessments (CFU and QSAR) of these IDPs revealed a notable reduction in toxicity relative to the parent compound. Additionally,

$Bi_4O_5I_2$@3DH maintained high photocatalytic performance in challenging water matrices (tap, pond, and secondary-treated wastewater) and in the presence of multiple cresol isomers. The composite also demonstrated good stability and reusability over successive cycles, highlighting its potential as a sustainable, robust, scalable, and field-applicable photocatalyst for water and wastewater remediation.

**References**

(1) Gupta, A. K.; Ayoob, S. *Fluoride in Drinking Water*; CRC Press, 2016. https://doi.org/10.1201/b21385.

(2) Rawat, A.; Srivastava, A.; Bhatnagar, A.; Gupta, A. K. Technological Advancements for the Treatment of Steel Industry Wastewater: Effluent Management and Sustainable Treatment Strategies. *J. Clean. Prod.* **2023**, *383* (November 2022), 135382. https://doi.org/10.1016/j.jclepro.2022.135382.

(3) Singh, A.; Srivastava, A.; Saidulu, D.; Gupta, A. K. Advancements of Sequencing Batch Reactor for Industrial Wastewater Treatment: Major Focus on Modifications, Critical Operational Parameters, and Future Perspectives. *J. Environ. Manage.* **2022**, *317*, 115305. https://doi.org/10.1016/j.jenvman.2022.115305.

(4) Rawat, A.; Srivastava, S. K.; Tiwary, C. S.; Gupta, A. K. Visible Light Driven Z-Scheme α-MnO2 (1D)/Bi7O9I3 (2D) Heterojunction Photocatalyst for Efficient Degradation of Bisphenol A in Water. *J. Environ. Chem. Eng.* **2024**, *12* (3), 112879. https://doi.org/10.1016/j.jece.2024.112879.

(5) Pillai, I. M. S.; Gupta, A. K. Batch and Continuous Flow Anodic Oxidation of 2,4-Dinitrophenol: Modeling, Degradation Pathway and Toxicity. *J. Electroanal. Chem.* **2015**, *756*, 108–117. https://doi.org/10.1016/j.jelechem.2015.08.020.

(6) Kennes, C.; Mendez, R.; Lema, J. M. Methanogenic Degradation of P-Cresol in Batch and in Continuous UASB Reactors. *Water Res.* **1997**, *31* (7), 1549–1554. https://doi.org/10.1016/S0043-1354(96)00156-X.

(7) Arya, D.; Kumar, S.; Kumar, S. Biodegradation Dynamics and Cell Maintenance for the Treatment of Resorcinol and P-Cresol by Filamentous Fungus Gliomastix Indicus. *J. Hazard. Mater.* **2011**, *198*, 49–56. https://doi.org/10.1016/j.jhazmat.2011.10.009.

(8) National Institute of Medical (NIM) DrugBank. *Cresols Profile*. https://pubchem.ncbi.nlm.nih.gov/compound/Cresol#section=Solubility (accessed 2024-12-06).

(9) Kumar, S.; Arya, D.; Malhotra, A.; Kumar, S.; Kumar, B. Biodegradation of Dual Phenolic Substrates in Simulated Wastewater by Gliomastix Indicus MTCC 3869. *J. Environ. Chem. Eng.* **2013**, *1* (4), 865–874. https://doi.org/10.1016/j.jece.2013.07.027.

(10) Zou, L.; Wang, Y.; Huang, C.; Li, B.; Lyu, J.; Wang, S.; Lu, H.; Li, J. Meta-Cresol Degradation by Persulfate through UV/O3 Synergistic Activation: Contribution of Free Radicals and Degradation Pathway. *Sci. Total Environ.* **2021**, *754*, 142219. https://doi.org/10.1016/j.scitotenv.2020.142219.

(11) Bera, S.; Kauser, H.; Mohanty, K. Optimization of P-Cresol Biodegradation Using Novel Bacterial Strains Isolated from Petroleum Hydrocarbon Fallout. *J. Water Process Eng.* **2019**, *31*, 100842. https://doi.org/10.1016/j.jwpe.2019.100842.

(12) Kuila, S. K.; Gorai, D. K.; Gupta, B.; Gupta, A. K.; Tiwary, C. S.; Kundu, T. K. Lanthanum Ions Decorated 2-Dimensional g-C3N4 for Ciprofloxacin Photodegradation. *Chemosphere* **2021**, *268*, 128780. https://doi.org/10.1016/j.chemosphere.2020.128780.

(13) Panigrahy, N.; Barik, M.; Sahoo, R. K.; Sahoo, N. K. Metabolic Profile Analysis and Kinetics of P-Cresol Biodegradation by an Indigenous Pseudomonas Citronellolis NS1 Isolated from Coke Oven Wastewater. *Int. Biodeterior. Biodegradation* **2020**, *147*, 104837. https://doi.org/10.1016/j.ibiod.2019.104837.

(14) USEPA. *Code of fedral regulations*. https://www.ecfr.gov/current/title-40/chapter-I/subchapter-D/part-116#116.4 (accessed 2024-12-06).

(15) Kumar, K. V. A.; Chandana, L.; Ghosal, P.; Subrahmanyam, C. Simultaneous Photocatalytic Degradation of p-Cresol and Cr (VI) by Metal Oxides Supported Reduced Graphene Oxide. *Mol. Catal.* **2018**, *451*, 87–95. https://doi.org/10.1016/j.mcat.2017.11.014.

(16) Melián, E. P.; Díaz, O. G.; Araña, J.; Rodríguez, J. M. D.; Rendón, E. T.; Melián, J. A. H. Kinetics and Adsorption Comparative Study on the Photocatalytic Degradation of O-, m- and p-Cresol. *Catal. Today* **2007**, *129* (1–2), 256–262. https://doi.org/10.1016/j.cattod.2007.08.003.

(17) Zhu, Y.; Kolar, P. Adsorptive Removal of P-Cresol Using Coconut Shell-Activated Char. *J. Environ. Chem. Eng.* **2014**, *2* (4), 2050–2058. https://doi.org/10.1016/j.jece.2014.08.022.

(18) Lee, K.-R.; Tan, C.-S. Separation of m- and p-Cresols in Compressed Propane Using Modified HZSM-5 Pellets. *Ind. Eng. Chem. Res.* **2000**, *39* (4), 1035–1038. https://doi.org/10.1021/ie990613o.

(19) Zhang, Y.; Zhuang, L.; Ji, B.; Ren, Y.; Xu, X.; He, J.; Xue, Y.; Sun, H. Ultrasonic Cavitation Treatment of O-Cresol Wastewater and Long-Term Pilot-Scale Study. *J. Environ. Manage.* **2025**, *375*, 124208. https://doi.org/10.1016/j.jenvman.2025.124208.

(20) Mahdavianpour, M.; Moussavi, G.; Farrokhi, M. Biodegradation and COD Removal of p-Cresol in a Denitrification Baffled Reactor: Performance Evaluation and Microbial Community. *Process Biochem.* **2018**, *69*, 153–160. https://doi.org/10.1016/j.procbio.2018.03.016.

(21) Feng, Y.; Li, H.; Ling, L.; Yan, S.; Pan, D.; Ge, H.; Li, H.; Bian, Z. Enhanced Photocatalytic Degradation Performance by Fluid-Induced Piezoelectric Field. *Environ. Sci. Technol.* **2018**, *52* (14), 7842–7848.

https://doi.org/10.1021/acs.est.8b00946.

(22) Lim, J.; Kim, H.; Park, J.; Moon, G.-H.; Vequizo, J. J. M.; Yamakata, A.; Lee, J.; Choi, W. How G-C 3 N 4 Works and Is Different from TiO 2 as an Environmental Photocatalyst: Mechanistic View. *Environ. Sci. Technol.* **2020**, *54* (1), 497–506. https://doi.org/10.1021/acs.est.9b05044.

(23) Wang, C.; Tian, J.; Tan, Q.; Xie, J.; Chen, G.; Chen, Y.; Mao, X. Uncovering Photocatalytic Mechanisms toward Water Treatment by Operando Super-Resolution Reaction Imaging. *Environ. Sci. Technol.* **2025**, *59* (20), 9865–9885. https://doi.org/10.1021/acs.est.5c00209.

(24) Dong, F.; Wang, Z.; Li, Y.; Ho, W.-K.; Lee, S. C. Immobilization of Polymeric G-C 3 N 4 on Structured Ceramic Foam for Efficient Visible Light Photocatalytic Air Purification with Real Indoor Illumination. *Environ. Sci. Technol.* **2014**, *48* (17), 10345–10353. https://doi.org/10.1021/es502290f.

(25) Rawat, A.; Srivastava, S. K.; Tiwary, C. S.; Gupta, A. K. An LED-Driven Hematite/Bi 4 O 5 I 2 Nanocomposite as an S-Scheme Heterojunction Photocatalyst for Efficient Degradation of Phenolic Compounds in Real Wastewater. *J. Mater. Chem. A* **2025**, *13* (2), 1271–1286. https://doi.org/10.1039/D4TA07324J.

(26) Parida, V. K.; Srivastava, S. K.; Chowdhury, S.; Gupta, A. K. Visible Light-Assisted Degradation of Sulfamethoxazole on 2D/0D Sulfur-Doped Bi 2 O 3 /MnO 2 Z-Scheme Heterojunction Immobilized Photocatalysts. *Langmuir* **2023**, *39* (51), 18846–18865. https://doi.org/10.1021/acs.langmuir.3c02733.

(27) Saidulu, D.; Bhatnagar, A.; Kumar Gupta, A. Integrated Anoxic/Oxic Biofilm Reactor and Photocatalytic Treatment for Enhanced Antibiotic Removal from Real Wastewater Matrices. *Sep. Purif. Technol.* **2024**, *350*, 128002.

https://doi.org/10.1016/j.seppur.2024.128002.

(28) Fu, G.-B.; Xie, R.; Qin, J.-W.; Deng, X.-B.; Ju, X.-J.; Wang, W.; Liu, Z.; Chu, L.-Y. Facile Fabrication of Photocatalyst-Immobilized Gel Beads with Interconnected Macropores for the Efficient Removal of Pollutants in Water. *Ind. Eng. Chem. Res.* **2021**, *60* (24), 8762–8775. https://doi.org/10.1021/acs.iecr.1c00971.

(29) Lim, D.-H.; Ali Maitlo, H.; A. Younis, S.; Kim, K.-H. The Practical Utility of Nitrogen Doped TiO2 as a Photocatalyst for the Oxidative Removal of Gaseous Formaldehyde. *Mater. Today Nano* **2024**, *27*, 100499. https://doi.org/10.1016/j.mtnano.2024.100499.

(30) Lisowski, P.; Colmenares, J. C.; Mašek, O.; Lisowski, W.; Lisovytskiy, D.; Kamińska, A.; Łomot, D. Correction for "Dual Functionality of TiO2/Biochar Hybrid Materials: Photocatalytic Phenol Degradation in the Liquid Phase and Selective Oxidation of Methanol in the Gas Phase." *ACS Sustain. Chem. Eng.* **2019**, *7* (19), 16933–16934. https://doi.org/10.1021/acssuschemeng.9b04559.

(31) Liu, J.; Du, T.; Chen, P.; Yue, Q.; Wang, H.; Zhou, L.; Wang, Y. Construction of Bi2WO6/g-C3N4/Cu Foam as 3D Z-Scheme Photocatalyst for Photocatalytic CO2 Reduction. *Appl. Surf. Sci.* **2024**, *664*, 160274. https://doi.org/10.1016/j.apsusc.2024.160274.

(32) Loeb, S. K.; Alvarez, P. J. J.; Brame, J. A.; Cates, E. L.; Choi, W.; Crittenden, J.; Dionysiou, D. D.; Li, Q.; Li-Puma, G.; Quan, X.; Sedlak, D. L.; David Waite, T.; Westerhoff, P.; Kim, J.-H. The Technology Horizon for Photocatalytic Water Treatment: Sunrise or Sunset? *Environ. Sci. Technol.* **2019**, *53* (6), 2937–2947. https://doi.org/10.1021/acs.est.8b05041.

(33) Willis, D. E.; Sheets, E. C.; Worbington, M. R.; Kamat, M.; Glass, S. K.; Caso, M. J.;

Ofoegbuna, T.; Diaz, L. M.; Osei-Appau, C.; Snow, S. D.; McPeak, K. M. Efficient Chemical-Free Degradation of Waterborne Micropollutants with an Immobilized Dual-Porous TiO 2 Photocatalyst. *ACS ES&T Eng.* **2023**, *3* (11), 1694–1705. https://doi.org/10.1021/acsestengg.3c00191.

(34) Guo, S.; Gao, X.; Huang, Y.; Zhou, R.; Chen, F.; Cai, C.; Zhou, K.; Chen, R. Efficient Degradation of Ciprofloxacin via Peroxymonosulfate Activation over a Hierarchically Porous Cu–Ti Alloy Manufactured by 3D Printing. *ACS ES&T Water* **2025**, *5* (1), 33–41. https://doi.org/10.1021/acsestwater.4c00383.

(35) Ghosal, P.; Gupta, B.; Ambekar, R. S.; Rahman, M. M.; Ajayan, P. M.; Aich, N.; Gupta, A. K.; Tiwary, C. S. 3D Printed Materials in Water Treatment Applications. *Adv. Sustain. Syst.* **2022**, *6* (3). https://doi.org/10.1002/adsu.202100282.

(36) Ambekar, R. S.; Joseph, A.; Ganji, S.; Agrawal, R.; Nirmal, G.; Tiwary, C. S. Printing Resolution Effect on Mechanical Properties of Porous Boehmite Direct Ink 3D Printed Structures. *Ceram. Int.* **2024**, *50* (21), 44447–44456. https://doi.org/10.1016/j.ceramint.2024.08.292.

(37) Roy Barman, S.; Gavit, P.; Chowdhury, S.; Chatterjee, K.; Nain, A. 3D-Printed Materials for Wastewater Treatment. *JACS Au* **2023**, *3* (11), 2930–2947. https://doi.org/10.1021/jacsau.3c00409.

(38) Das, R.; de Oliveira, R. B.; Kumar, B.; Mishra, V.; Sarkar, S.; Sarkar, S.; Felix, I. de M.; Machado, L. D.; Tiwary, C. S. Engineering the Atomic Interface of Refractory-Metal-Reinforced Copper Matrix Using Direct Ink 3D Printing. *Adv. Eng. Mater.* **2025**, *27* (2). https://doi.org/10.1002/adem.202401747.

(39) Saadi, M. A. S. R.; Maguire, A.; Pottackal, N. T.; Thakur, M. S. H.; Ikram, M. M.; Hart, A. J.; Ajayan, P. M.; Rahman, M. M. Direct Ink Writing: A 3D Printing

Technology for Diverse Materials. *Adv. Mater.* **2022**, *34* (28), 2108855. https://doi.org/10.1002/adma.202108855.

(40) Majooni, Y.; Abioye, S. O.; Fayazbakhsh, K.; Yousefi, N. Nano-Enabled 3D-Printed Structures for Water Treatment. *ACS ES&T Water* **2024**, *4* (5), 1952–1965. https://doi.org/10.1021/acsestwater.3c00770.

(41) Wei, P.; Cipriani, C. E.; Pentzer, E. B. Thermal Energy Regulation with 3D Printed Polymer-Phase Change Material Composites. *Matter* **2021**, *4* (6), 1975–1989. https://doi.org/10.1016/j.matt.2021.03.019.

(42) Wei, P.; Leng, H.; Chen, Q.; Advincula, R. C.; Pentzer, E. B. Reprocessable 3D-Printed Conductive Elastomeric Composite Foams for Strain and Gas Sensing. *ACS Appl. Polym. Mater.* **2019**, *1* (4), 885–892. https://doi.org/10.1021/acsapm.9b00118.

(43) Chen, Q.; Cao, P.; Advincula, R. C. Mechanically Robust, Ultraelastic Hierarchical Foam with Tunable Properties via 3D Printing. *Adv. Funct. Mater.* **2018**, *28* (21). https://doi.org/10.1002/adfm.201800631.

(44) Ambekar, R. S.; Joseph, A.; Ganji, S.; Nirmal, G.; Agrawal, R.; Tiwary, C. S. Enhanced Mechanical Properties of Direct Ink Writing (3D Printed) Hexagonal Boron Nitride Reinforced Porous Boehmite Structures. *Adv. Eng. Mater.* **2024**, *26* (7). https://doi.org/10.1002/adem.202301830.

(45) Wang, L.; Zhou, J.-C.; Li, Z.-H.; Zhang, X.; Leung, K. M. Y.; Yuan, L.; Sheng, G.-P. Facet-Specific Photocatalytic Degradation of Extracellular Antibiotic Resistance Genes by Hematite Nanoparticles in Aquatic Environments. *Environ. Sci. Technol.* **2023**, *57* (51), 21835–21845. https://doi.org/10.1021/acs.est.3c06571.

(46) Xiao, Z.; Xiao, J.; Yuan, L.; Ai, M.; Idrees, F.; Huang, Z.-F.; Shi, C.; Zhang, X.; Pan,

L.; Zou, J.-J. Z-Scheme Charge Transfer between a Conjugated Polymer and α-Fe 2 O 3 for Simultaneous Photocatalytic H 2 Evolution and Ofloxacin Degradation. *J. Mater. Chem. A* **2024**, *12* (9), 5366–5376. https://doi.org/10.1039/D3TA07217G.

(47) Rasool, A. T.; Nazir, A.; Zhang, Q.; Li, E. Fabrication of Novel TiN@Cu2O Nanocomposite for Efficient Photodegradation of Sulfamethoxazole by Peroxymonosulfate Stimulation and Bacterial Inactivation: DFT Analysis and Mechanism Insight. *J. Environ. Chem. Eng.* **2025**, *13* (5), 117750. https://doi.org/10.1016/j.jece.2025.117750.

(48) CCTE, E. Toxicity Estimation Software Tool (TEST). The United States Environmental Protection Agency's Center for Computational Toxicology and Exposure. Software. 2022. https://doi.org/https://doi.org/10.23645/epacomptox.21379365.v3.

(49) Nath, S. S.; Patil, I. G.; Sundriyal, P. Material Extrusion of Electrochemical Energy Storage Devices for Flexible and Wearable Electronic Applications. *J. Energy Storage* **2024**, *79*, 110129. https://doi.org/10.1016/j.est.2023.110129.

(50) Moreno-Sanabria, L.; Ramírez, C.; Osendi, M. I.; Belmonte, M.; Miranzo, P. Enhanced Thermal and Mechanical Properties of 3D Printed Highly Porous Structures Based on Γ-Al 2 O 3 by Adding Graphene Nanoplatelets. *Adv. Mater. Technol.* **2022**, *7* (9). https://doi.org/10.1002/admt.202101455.

(51) Ambekar, R. S.; Joseph, A.; Ganji, S.; Nirmal, G.; Agrawal, R.; Tiwary, C. S. Enhanced Mechanical Properties of Direct Ink Writing ( 3D Printed ) Hexagonal Boron Nitride Reinforced Porous Boehmite Structures. *Adv. Eng. Mater.* **2024**, 2301830. https://doi.org/10.1002/adem.202301830.

(52) Zhang, D.; Chu, C.; Ma, S.; Wang, Y.; Duan, C.; Guo, J.; Shi, X.; Xu, G.; Cheng, Y.;

Sun, A. A Novel Method to Avoid the Sintering Shrinkage of Al2O3-Cr Cermets Formed by Direct Ink Writing. *J. Alloys Compd.* **2023**, *931*, 167632. https://doi.org/10.1016/j.jallcom.2022.167632.

(53) Ye, Z.; Chu, C.; Zhang, D.; Ma, S.; Guo, J.; Cheng, Y.; Xu, G.; Li, Z.; Sun, A. Study on 3D-Direct Ink Writing Based on Adding Silica Submicron-Particles to Improve the Rheological Properties of Alumina Ceramic Ink. *Mater. Today Commun.* **2021**, *28*, 102534. https://doi.org/10.1016/j.mtcomm.2021.102534.

(54) Lewis, J. A.; Smay, J. E.; Stuecker, J.; Cesarano, J. Direct Ink Writing of Three-Dimensional Ceramic Structures. *J. Am. Ceram. Soc.* **2006**, *89* (12), 3599–3609. https://doi.org/10.1111/j.1551-2916.2006.01382.x.

(55) Neumann, T. V.; Dickey, M. D. Liquid Metal Direct Write and 3D Printing: A Review. *Adv. Mater. Technol.* **2020**, *5* (9). https://doi.org/10.1002/admt.202000070.

(56) Haile, B. S.; Pal, V.; Pal, T.; Slathia, S.; Jigi, G. M.; Negedu, S. D.; Tiwari, N.; Singh, H.; Joseph, A.; Olu, F. E.; Tiwary, C. S. Direct Ink Writing (3D Printing) of Robust, Highly Efficient, Double-Half-Heusler Thermoelectric High-Entropy Alloy. *Adv. Eng. Mater.* **2025**, *27* (7). https://doi.org/10.1002/adem.202402283.

(57) Moradlou, O.; Rabiei, Z.; Banazadeh, A.; Warzywoda, J.; Zirak, M. Carbon Quantum Dots as Nano-Scaffolds for α-Fe2O3 Growth: Preparation of Ti/CQD@α-Fe2O3 Photoanode for Water Splitting under Visible Light Irradiation. *Appl. Catal. B Environ.* **2018**, *227*, 178–189. https://doi.org/10.1016/j.apcatb.2018.01.016.

(58) R. L. Blake, R. E. Hessevick, Tibor Zoltai, L. W. F. Refinement of the Hematite Structure. *Am. Mineral. J. Earth Planet. Mater.* **1966**, *51 (1-2)*, 123–129.

(59) Xiao, X.; Xing, C.; He, G.; Zuo, X.; Nan, J.; Wang, L. Solvothermal Synthesis of

Novel Hierarchical Bi4O5I2 Nanoflakes with Highly Visible Light Photocatalytic Performance for the Degradation of 4-Tert-Butylphenol. *Appl. Catal. B Environ.* **2014**, *148–149*, 154–163. https://doi.org/10.1016/j.apcatb.2013.10.055.

(60) Keller, E.; Krämer, V.; Schmidt, M.; Oppermann, H. The Crystal Structure of Bi 4 O 5 I 2 and Its Relation to the Structure of Bi 4 O 5 Br 2. *Zeitschrift für Krist. - Cryst. Mater.* **2002**, *217* (6), 256–264. https://doi.org/10.1524/zkri.217.6.256.22811.

(61) Blake, R. L.; HessevicK, R. E.; Zoltai, T.; Finger, L. W. Refinement of the Hematite Structure. *Am. Mineral.* **1966**, *51* (Three-dimensional dilTraction intensities u.ere collected from a spherical single crystal of hematite, a FezOs, with a Buerger single-crystal diffractometer. The structure has been refined with a least-squares program, and the final structure gave an R-fa), 123–129.

(62) Mashkoor, F.; Shoeb, M.; Zhu, S.; Jeong, H.; Jeong, C. Effect of MXene on Co 3 O 4 –LaVO 4 Nanocomposites for Synergistic Charge Transport Enhancement and High-Performance VARTM Assisted Solid-State Supercapacitor Devices Using Woven Carbon Fiber. *J. Mater. Chem. A* **2025**. https://doi.org/10.1039/D5TA02046H.

(63) Chen, Y.; Nisar, M.; Qin, W.; Xu, Z.; Danish, M. H.; Li, F.; Liang, G.; Ge, Z.; Luo, J.; Zheng, Z. Integration of Boron Nitride Into Tin-Enriched SnSe 2 for a High-performance Thermoelectric Nanocomposite with Optimized Electrical Transport and Mechanical Properties. *Adv. Funct. Mater.* **2025**. https://doi.org/10.1002/adfm.202425050.

(64) khan, I.; Sun, Y.; Khan, F.; Zhang, J.; Kareem, A.; Naseem, M.; Ali, Z.; Sultan, M.; Arif, U.; Ma, X.; Wu, Z. Synthesis, Mechanism and Environmental Applications of g-C3N4 Composites: A Synergistic Approach to Adsorption and Photocatalysis. *Sep. Purif. Technol.* **2025**, *359*, 130472. https://doi.org/10.1016/j.seppur.2024.130472.

(65) Chen, Y.; Dionysiou, D. D. Correlation of Structural Properties and Film Thickness to Photocatalytic Activity of Thick TiO2 Films Coated on Stainless Steel. *Appl. Catal. B Environ.* **2006**, *69* (1–2), 24–33. https://doi.org/10.1016/j.apcatb.2006.05.002.

(66) Zakria, H. S.; Othman, M. H. D.; Kamaludin, R.; Sheikh Abdul Kadir, S. H.; Kurniawan, T. A.; Jilani, A. Immobilization Techniques of a Photocatalyst into and onto a Polymer Membrane for Photocatalytic Activity. *RSC Adv.* **2021**, *11* (12), 6985–7014. https://doi.org/10.1039/d0ra10964a.

(67) Claes, T.; Gerven, T. Van; Leblebici, M. E. Design Considerations for Photocatalytic Structured Packed Bed Reactors. *Chem. Eng. J.* **2021**, *403*, 126355. https://doi.org/10.1016/j.cej.2020.126355.

(68) Gupta, B.; Saidulu, D.; Srivastava, A.; Rawat, A.; Singh, A.; Bhatnagar, A.; Gupta, A. K. Performance Assessment of Photo-Controlled Catalytic and Moving Biocarrier-Based Hybrid Systems for the Treatment of Pharmaceuticals from Municipal and Industrial Wastewater: Insights into Influencing Factors and Degradation Mechanisms. *J. Environ. Chem. Eng.* **2023**, *11* (3), 110128. https://doi.org/10.1016/j.jece.2023.110128.

(69) Li, R.; Song, X.; Huang, Y.; Fang, Y.; Jia, M.; Ma, W. Visible-Light Photocatalytic Degradation of Azo Dyes in Water by Ag3PO4: An Unusual Dependency between Adsorption and the Degradation Rate on PH Value. *J. Mol. Catal. A Chem.* **2016**, *421*, 57–65. https://doi.org/10.1016/j.molcata.2016.05.009.

(70) National Center for Biotechnology Information. *PubChem Compound Summary for CID 2879, P-Cresol*. https://pubchem.ncbi.nlm.nih.gov/compound/P-Cresol.

(71) Pearce, P. J.; Simkins, R. J. J. Acid Strengths of Some Substituted Picric Acids. *Can. J. Chem.* **1968**, *46* (2), 241–248. https://doi.org/10.1139/v68-038.

(72) Malefane, M. E. Co 3 O 4 /Bi 4 O 5 I 2 /Bi 5 O 7 I C-Scheme Heterojunction for Degradation of Organic Pollutants by Light-Emitting Diode Irradiation. *ACS Omega* **2020**, *5* (41), 26829–26844. https://doi.org/10.1021/acsomega.0c03881.

(73) Xiong, X.; Zhang, X.; Liu, S.; Zhao, J.; Xu, Y. Sustained Production of H2O2 in Alkaline Water Solution Using Borate and Phosphate-Modified Au/TiC2 Photocatalysts. *Photochem. Photobiol. Sci.* **2018**, *17* (8), 1018–1022. https://doi.org/10.1039/c8pp00177d.

(74) Zhu, C.; Wang, Y.; Qiu, L.; Liu, Y.; Li, H.; Yu, Y.; Li, J.; Yang, W. 3D Hierarchical Fe-Doped Bi4O5I2 Microflowers as an Efficient Fenton Photocatalyst for Tetracycline Degradation over a Wide PH Range. *Sep. Purif. Technol.* **2022**, *290*, 120878. https://doi.org/10.1016/j.seppur.2022.120878.

(75) Lai, C.; Huang, F.; Zeng, G.; Huang, D.; Qin, L.; Cheng, M.; Zhang, C.; Li, B.; Yi, H.; Liu, S.; Li, L.; Chen, L. Fabrication of Novel Magnetic MnFe2O4/Bio-Char Composite and Heterogeneous Photo-Fenton Degradation of Tetracycline in near Neutral PH. *Chemosphere* **2019**, *224*, 910–921. https://doi.org/10.1016/j.chemosphere.2019.02.193.

(76) Yin, R.; Li, Y.; Zhong, K.; Yao, H.; Zhang, Y.; Lai, K. Multifunctional Property Exploration: Bi 4 O 5 I 2 with High Visible Light Photocatalytic Performance and a Large Nonlinear Optical Effect. *RSC Adv.* **2019**, *9* (8), 4539–4544. https://doi.org/10.1039/C8RA08984A.

(77) Naveas, N.; Pulido, R.; Marini, C.; Hernández-Montelongo, J.; Silván, M. M. First-Principles Calculations of Hematite (α-Fe2O3) by Self-Consistent DFT+U+V. *iScience* **2023**, *26* (2), 106033. https://doi.org/10.1016/j.isci.2023.106033.

(78) Li, J.; Chu, D. Energy Band Engineering of Metal Oxide for Enhanced Visible Light

Absorption. In *Multifunctional Photocatalytic Materials for Energy*; Elsevier, 2018; pp 49–78. https://doi.org/10.1016/B978-0-08-101977-1.00005-3.

(79) Padilha, A. C. M.; Soares, M.; Leite, E. R.; Fazzio, A. Theoretical and Experimental Investigation of 2D Hematite. *J. Phys. Chem. C* **2019**, *123* (26), 16359–16365. https://doi.org/10.1021/acs.jpcc.9b01046.

(80) Jain, M.; Chelikowsky, J. R.; Louie, S. G. Reliability of Hybrid Functionals in Predicting Band Gaps. *Phys. Rev. Lett.* **2011**, *107* (21), 216806. https://doi.org/10.1103/PhysRevLett.107.216806.

(81) Ojha, N.; Pant, K. K.; Coy, E. Photocatalytic Conversion of Carbon Dioxide and Nitrogen Dioxide: Current Developments, Challenges, and Perspectives. *Ind. Eng. Chem. Res.* **2023**, *62* (51), 21885–21908. https://doi.org/10.1021/acs.iecr.3c03426.

(82) Deng, H.; Fei, X.; Yang, Y.; Fan, J.; Yu, J.; Cheng, B.; Zhang, L. S-Scheme Heterojunction Based on p-Type ZnMn2O4 and n-Type ZnO with Improved Photocatalytic CO2 Reduction Activity. *Chem. Eng. J.* **2021**, *409*, 127377. https://doi.org/10.1016/j.cej.2020.127377.

(83) Yang, H.; Wang, R.; Hou, H.; Li, Y.; Li, Z.; Wang, H.; Zhan, X.; Zhang, D.; Liang, Z.; Luo, Y.; Yang, W. Dual-Driven Charge Transport Enabled by S-Scheme Heterojunction and Solid Solution in CdS@N-NiCoO Photocatalysts for Enhanced Hydrogen Evolution. *Sep. Purif. Technol.* **2025**, *362*, 131819. https://doi.org/10.1016/j.seppur.2025.131819.

(84) Khunphonoi, R.; Grisdanurak, N. Mechanism Pathway and Kinetics of p -Cresol Photocatalytic Degradation over Titania Nanorods under UV − Visible Irradiation. *Chem. Eng. J.* **2016**, *296*, 420–427. https://doi.org/10.1016/j.cej.2016.03.117.

(85) GHS. *Globally Harmonized System of Classification and Labeling of Chemicals (GHS)*; 2021. https://unece.org/transport/standards/transport/dangerous-goods/ghs-rev9-2021.

(86) Parida, V. K.; Srivastava, S. K.; Gupta, A. K.; Rawat, A. A Review on Nanomaterial-Based Heterogeneous Photocatalysts for Removal of Contaminants from Water. **2023**, 1–38. https://doi.org/10.1166/mex.2023.2319.

(87) Rao, K. V. S.; Subrahmanyam, M.; Boule, P. Immobilized TiO2 Photocatalyst during Long-Term Use: Decrease of Its Activity. *Appl. Catal. B Environ.* **2004**, *49* (4), 239–249. https://doi.org/10.1016/j.apcatb.2003.12.017.

# List of Figures

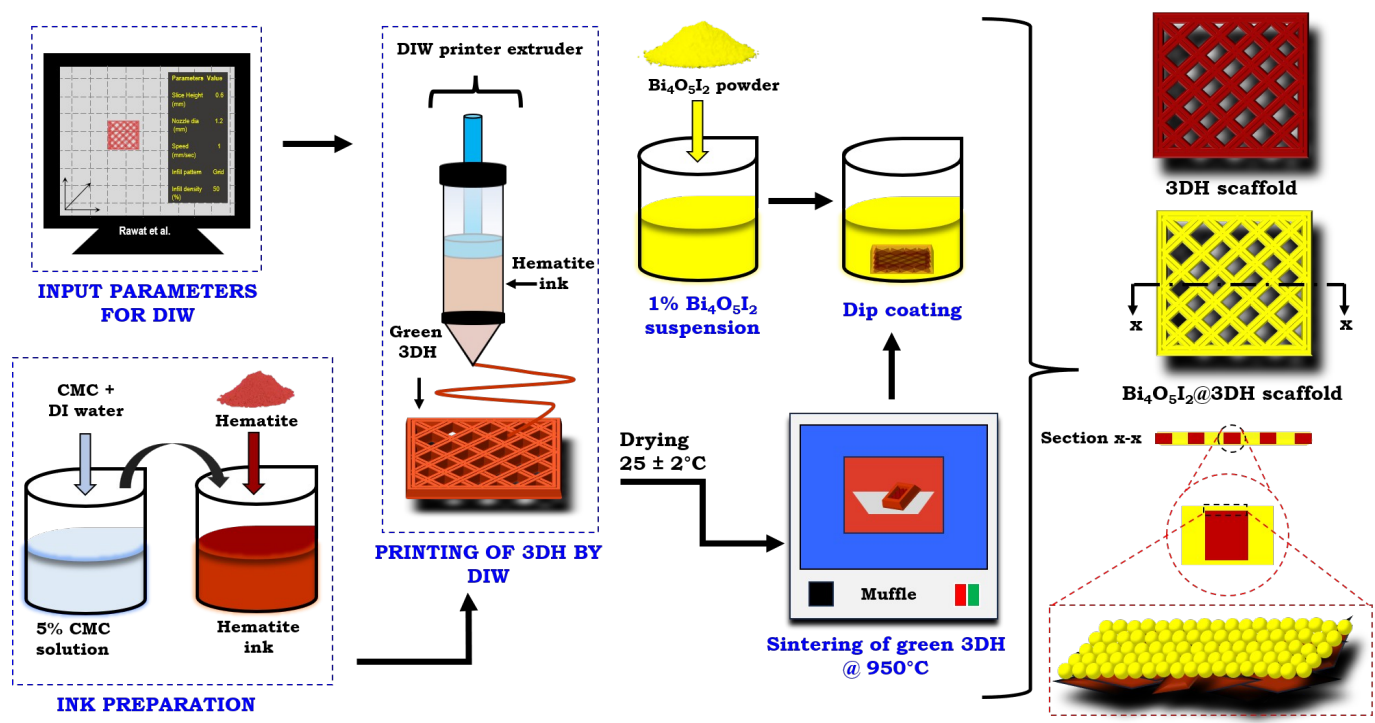

**Fig. 1.** Fabrication of $Bi_4O_5I_2$ decorated hematite scaffold ($Bi_4O_5I_2$@3DH) using the REPETREL software to define hematite printing parameters for direct ink writing, followed by dip-coating for immobilization of $Bi_4O_5I_2$ nanoparticles onto the sintered hematite grid.

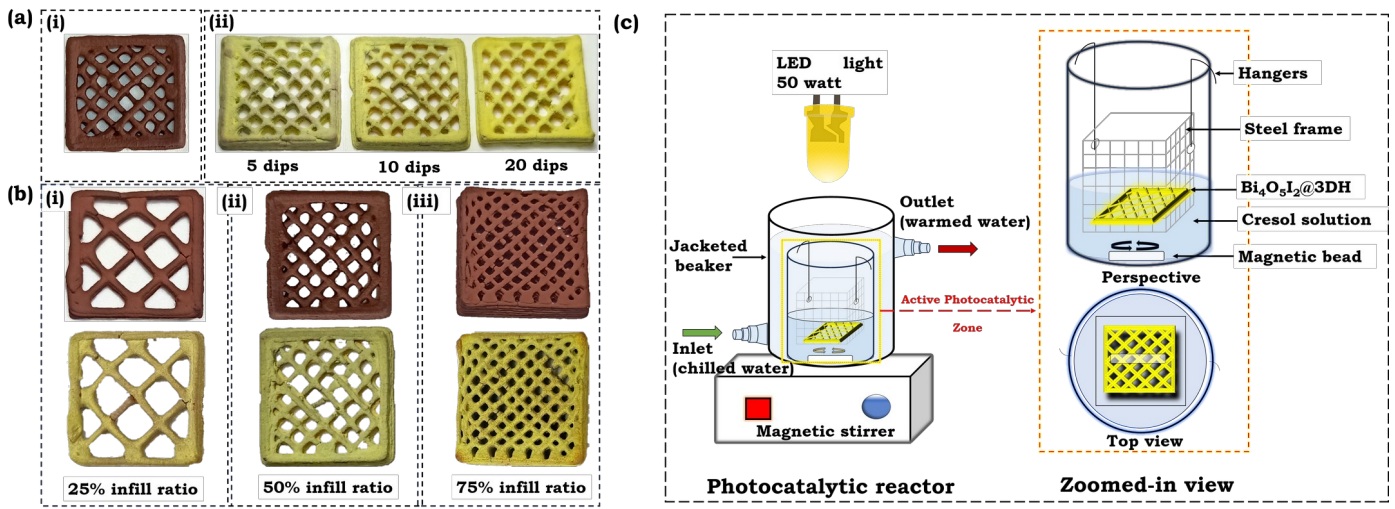

**Fig. 2.** (a) Variation in the number of dip cycles (N) for 3D-Printed hematite scaffold (i), decorated with $Bi_4O_5I_2$ (i.e., $Bi_4O_5I_2$@3DH) for (ii) N = 5, 10, and 20. (b) Bare hematite and $Bi_4O_5I_2$@3DH (N = 10) with varying grid infill density ratios, (i) 25%, (ii) 50%, and (iii) 75%. (c) Experimental set-up endowed with the zoomed-in view, employed for the photocatalysis experiment.

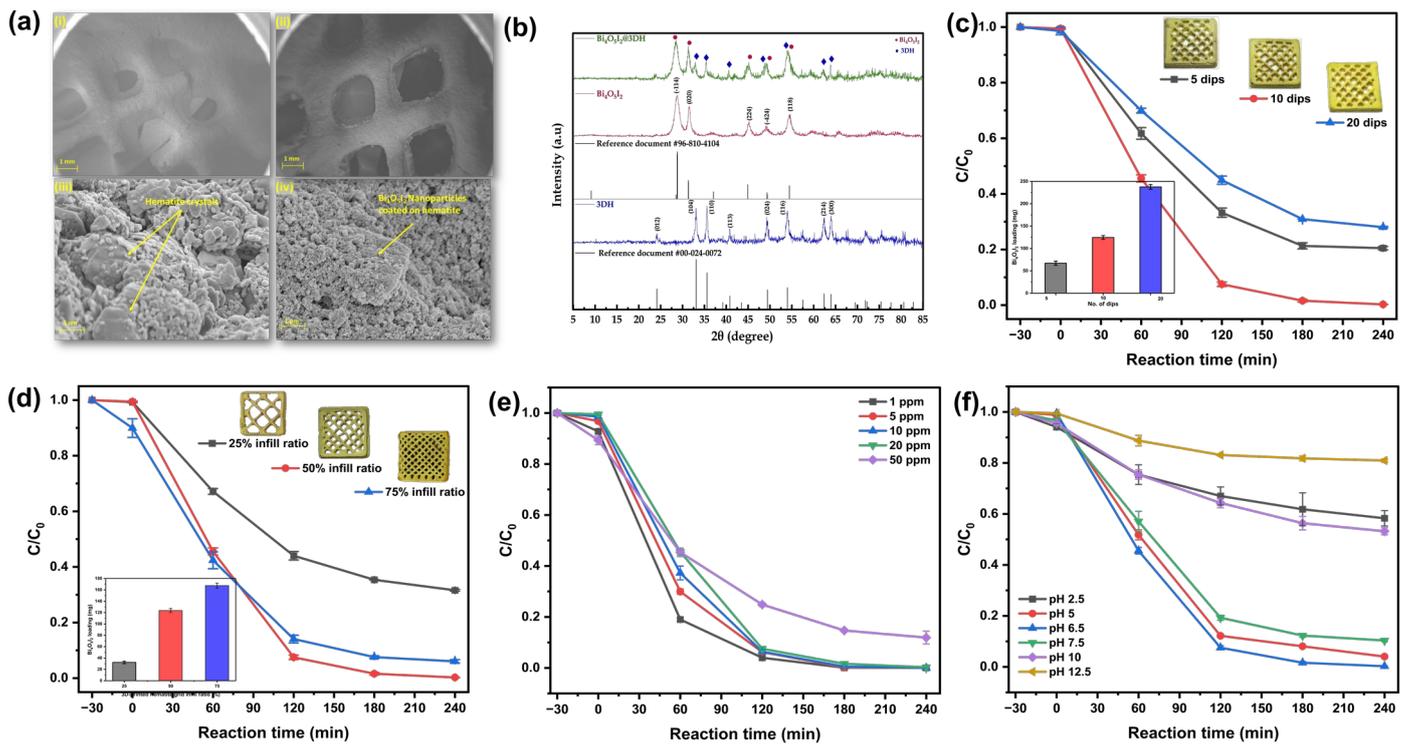

**Fig. 3.** (a) FEG-SEM images of materials: low magnification SEM images of (i) 3D-printed hematite, i.e., 3DH, and (ii) $Bi_4O_5I_2$-coated 3DH, i.e., $Bi_4O_5I_2$@3DH; high magnification SEM images of (iii) 3DH and (iv) $Bi_4O_5I_2$@3DH. (b) XRD pattern of $Bi_4O_5I_2$@3DH, $Bi_4O_5I_2$, reference document no. 96-810-4104, 3DH, and reference document no. 00-024-0072. Influence of operational parameters on the photocatalytic degradation of model contaminant (*p*-cresol): (c) number of $Bi_4O_5I_2$ solution dips (N = 5, 10, and 20) [provided: 50% grid IDR, initial *p*-cresol concentration ($C_0$) = 20 ppm, and pH = 6.5], (d) IDR of 3DH (= 25, 50, and 75%) [provided: N=10, $C_0$ = 20 ppm, and pH = 6.5], (e) C0 [provided: N = 10, IDR = 50%, and pH = 6.5] and (f) initial pH of p-cresol solution, [provided: N = 10, IDR = 50%, and C0 = 20 mg/L].

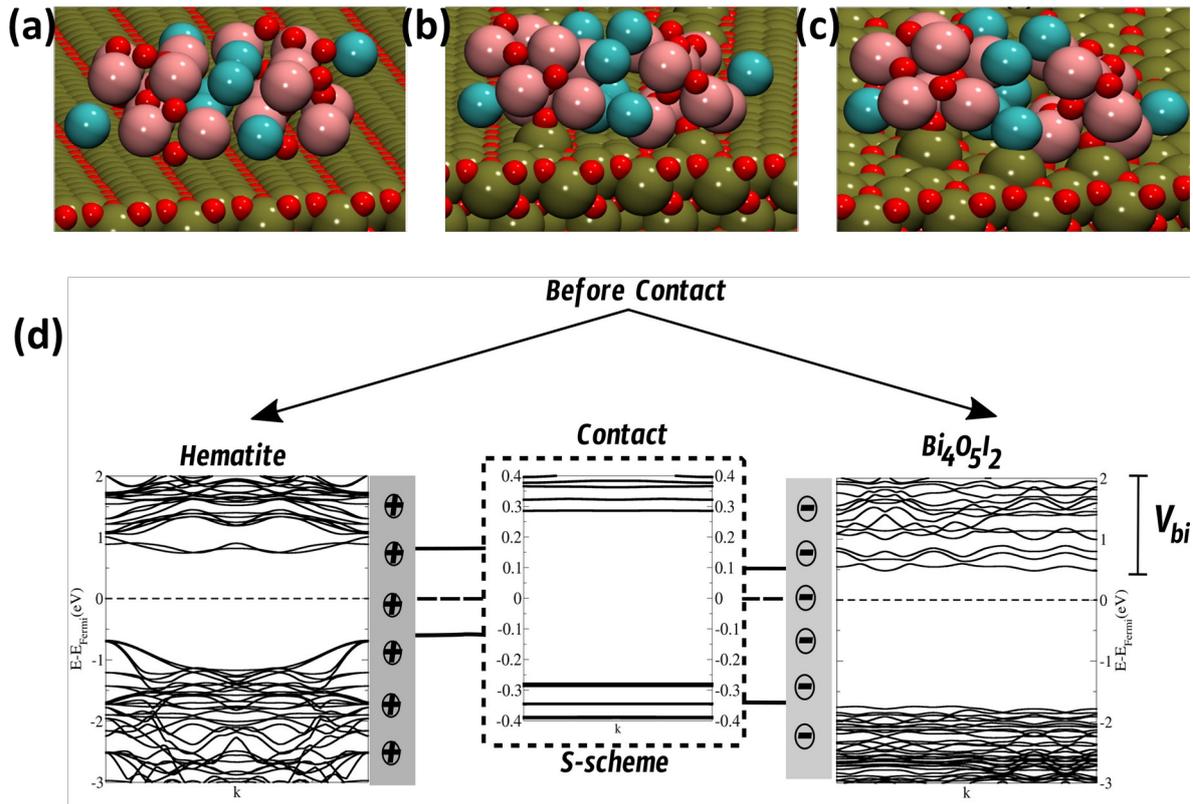

**Fig. 4.** Deposition of $Bi_4O_5I_2$ on hematite (001): (a) initial configuration with the flake positioned above the substrate; (b) intermediate state during optimization; (c) final structure showing interfacial Fe-O. (d) Band structure evolution of the $Bi_4O_5I_2$/hematite heterojunction.

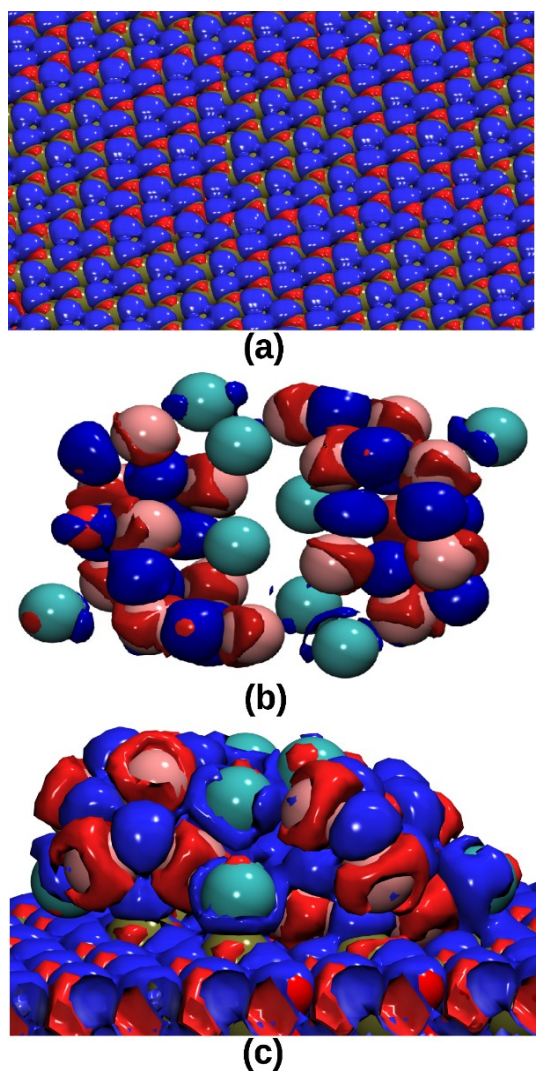

**Fig. 5**. Charge density difference (Δρ) maps. (a) Top view of the hematite (001) monolayer; (b) Isolated $Bi_4O_5I_2$ flake; (c) side view of the hematite/$Bi_4O_5I_2$ heterostructure interface. Red and blue isosurfaces represent regions of electron depletion (Δρ < 0) and accumulation (Δρ > 0), respectively, highlighting the interfacial charge transfer.

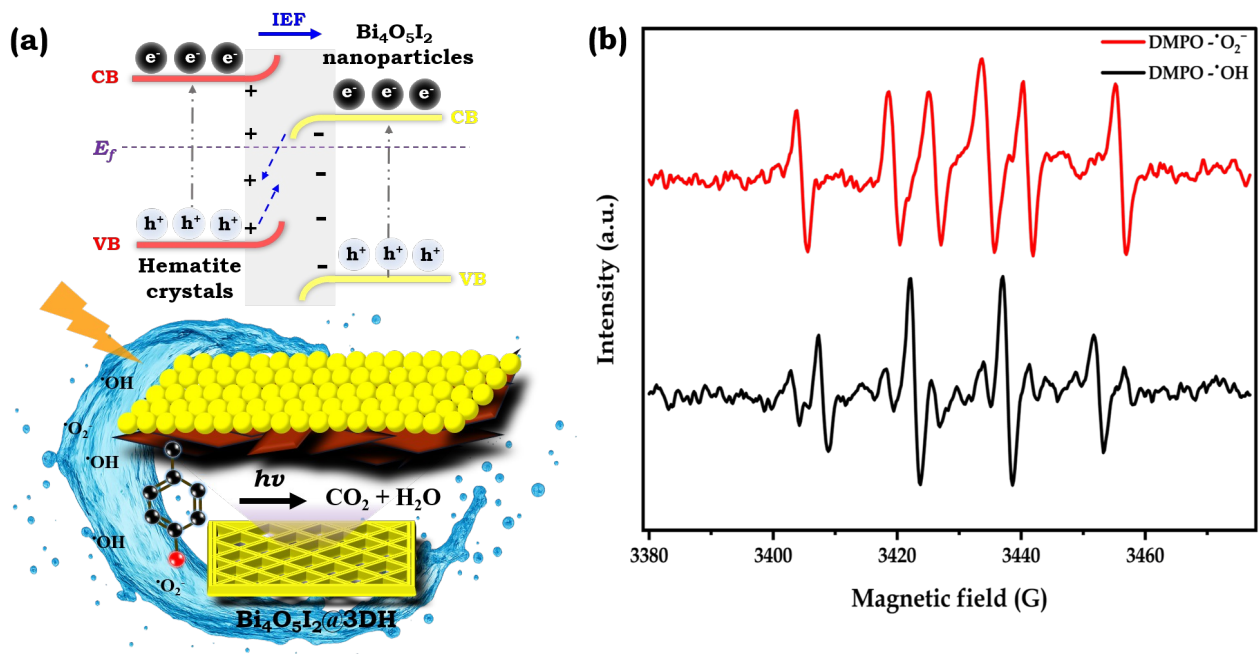

**Fig. 6.** (a) Electrons and holes transfer mechanism at the heterointerface of hematite and Bi$_4$O$_5$I$_2$ photocatalysts. (b) Electron paramagnetic resonance spectra of DMPO-·O$_2^-$ and DMPO-·OH adduct signals after 20 min irradiation.

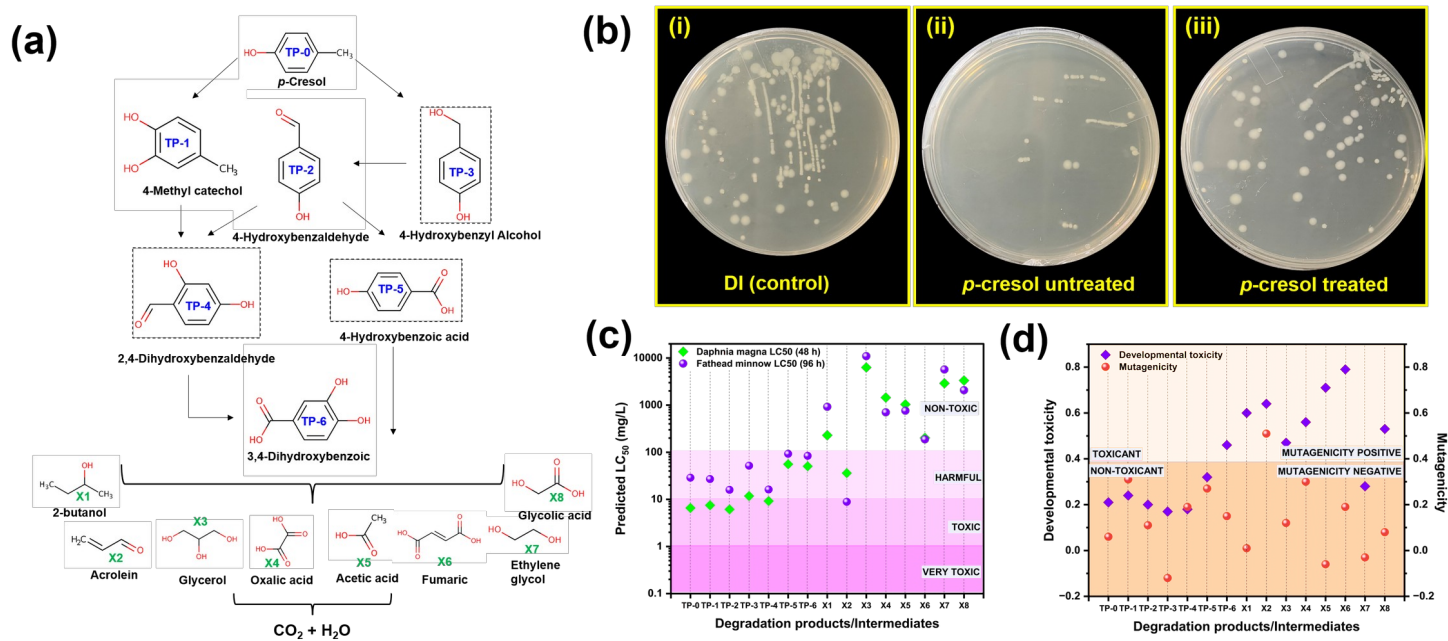

**Fig. 7.** (a) Plausible degradation pathways of *p*-cresol. (b) *In vitro* (CFU) bio-toxicity test: (i) DI water, i.e., control, (ii) untreated p-cresol sample, and (iii) 240 min photocatalytically degraded *p*-cresol sample. QSAR model-based TEST software profiles: (c)* predicted $LC_{50}$ values for *Daphnia magna* (48 h) and *Fathead minnow* (96 h), and (d)⊤ binary endpoints, i.e., developmental toxicity and mutagenicity. [*Very Toxic (≤ 1 mg/L), Toxic (2–10 mg/L), Harmful (11–100 mg/L), and non-toxic (≥ 100 mg/L).[85] ⊤If calculated score < 0.5, then activity = negative else if calculated score ≥ 0.5, then activity = positive[85]].

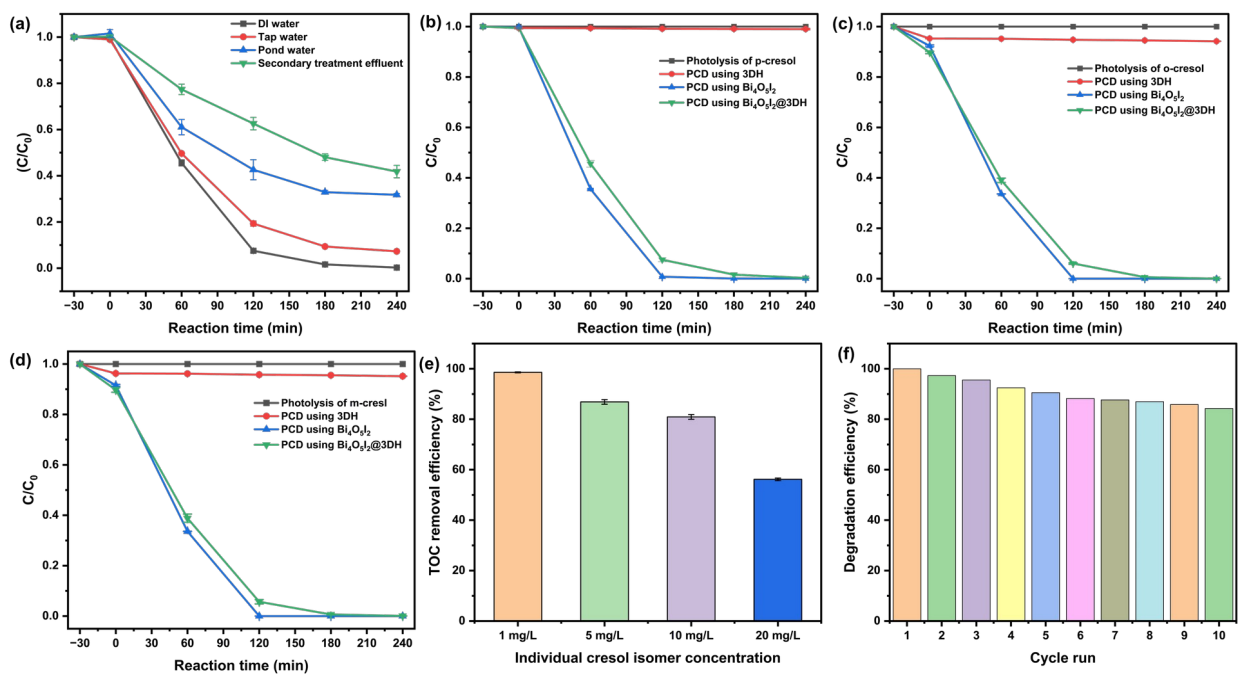

**Fig. 8.** Practical significance of Bi4O5I2@3DH composite: (a) Application of Bi$_4$O$_5$I$_2$@3DH in various water matrices, (b) the individual PCD performance of o-cresol and m-cresol. (c) Simultaneous degradation of the isomeric mixture was evaluated in terms of TOC removal efficiency and (d) TREs for *o*-cresol and *m*-cresol. (e) Reusability of as-fabricated Bi$_4$O$_5$I$_2$@3DH scaffold over 10 cycles.